\documentstyle[jkas2]{article} 
\runningauthor{KANG} 
\runningtitle{Cosmic Ray Spectrum in SNRs} 
\beginpage{1} 
\endpage{15} 

\def\etal{{\it et al.~}}
\def\eg{{\it e.g.,~}}
\def\ie{{\it i.e.,~}}
\def\kms{~{\rm km~s^{-1}}}
\def\cm3{~{\rm cm^{-3}}}

\def\lsim{\mathrel{  
        \raise0.3ex\hbox{$<$}\kern-0.75em{\lower0.65ex\hbox{$\sim$}}}}
\def\gsim{\mathrel{
        \raise0.3ex\hbox{$>$}\kern-0.75em{\lower0.65ex\hbox{$\sim$}}}}

\begin{document}

\title{COSMIC RAY SPECTRUM IN SUPERNOVA REMNANT SHOCKS}

\author{HYESUNG KANG}

\address{Department of Earth Sciences, Pusan National University, Pusan  609
-735, Korea \\
{\it E-mail: hskang@pusan.ac.kr }}

\address{\normalsize{\it (Received February 19, 2010; Accepted March 15, 2010)}}

\abstract{
We perform kinetic simulations of diffusive shock acceleration (DSA) in Type Ia 
supernova remnants (SNRs) expanding into a uniform interstellar medium (ISM).
Bohm-like diffusion due to self-excited Alfv\'en waves is assumed,
and simple models for Alfv\'enic drift and dissipation are adopted. 
Phenomenological models for thermal leakage injection are considered as well. 
We find that the preshock gas temperature is the primary parameter that
governs the cosmic ray (CR) acceleration efficiency and energy spectrum,
while the CR injection rate is a secondary parameter. 
For SNRs in the warm ISM of $T_0 \lsim 10^5$K, if the injection fraction is
$\xi \gsim 10^{-4}$, 
the DSA is efficient enough to convert more than 20 \% of the SN explosion
energy into CRs and the accelerated CR spectrum exhibits a concave curvature
flattening to $E^{-1.6}$, which is characteristic of CR modified shocks.
Such a flat source spectrum near the knee energy, however, may not be reconciled 
with the CR spectrum observed at Earth.
On the other hand, SNRs in the hot ISM of $T_0\approx 10^6$K 
with a small injection fraction, $\xi < 10^{-4}$, are 
inefficient accelerators with less than 10 \% of the explosion energy getting converted to CRs. 
Also the shock structure is almost test-particle like and the ensuing CR spectrum 
can be steeper than $E^{-2}$. 
With amplified magnetic field strength of order of 30$\mu$G, Alfv\'en
waves generated by the streaming instability may drift upstream fast enough 
to make the modified test-particle power-law as steep as $E^{-2.3}$, 
which is more consistent with the observed CR spectrum. 
}

\keywords{cosmic ray acceleration -- supernova remnants --  
hydrodynamics -- methods:numerical}
\maketitle

\section{INTRODUCTION}

It is believed that most of the Galactic cosmic rays (CRs) are accelerated in
the blast waves driven by supernova (SN) explosions (\eg Blandford \& Eichler 1987,
Reynolds 2008 and references therein).
If about 10 \% of Galactic SN luminosity, 
$L_{SN}\approx 10^{42} {\rm erg~s^{-1}}$, is transfered to the CR component,
the diffusive shock acceleration (DSA) at supernova remnants (SNRs) can
provide the CR luminosity, $L_{CR} \approx 10^{41} {\rm erg~s^{-1}}$
that escapes from the Galaxy.
Several time-dependent, kinetic simulations of the CR acceleration at SNRs
have shown that an order of 10 \% of the SN explosion energy can be converted 
to CRs, 
when a fraction $\sim 10^{-4}$ of incoming thermal particles are injected
into the CR population at the subshock 
(\eg Berezhko, \& V\"olk 1997;  Berezhko et al 2003; Kang 2006).

X-ray observations of young SNRs such as SN1006 and RCW86 
indicate the presence of 10-100 TeV electrons emitting nonthermal
synchrotron emission immediately inside the outer SNR shock
(Koyama \etal 1995; Bamba \etal 2006, Helder \etal 2009).
They provide clear evidence for the efficient acceleration of the CR electrons 
at SNR shocks.   
Moreover, HESS gamma-ray telescope detected TeV emission from several SNRs such as RXJ1713.7-3946, 
Cas A, Vela Junior, and RCW86, which may indicate possible detection of $\pi^0$ $\gamma$- 
rays produced by nuclear collisions of hadronic CRs with the surrounding gas
(Aharonian \etal 2004, 2009; Berezhko \& V\"olk 2006; Berezhko \etal 2009; Morlino \etal 2009,
Abdo \etal 2010). 
It is still challenging to discern whether such emission could provide direct evidence 
for the acceleration of hadronic CRs, since $\gamma$-ray emission could be produced by inverse Compton
scattering of the background radiation by X-ray emitting relativistic electrons.  
More recently, however, Fermi LAT has observed in GeV range several SNRs interacting with 
molecular clouds, providing some very convincing evidence of $\pi^0$ decay $\gamma$-rays 
(Abdo \etal 2009, 2010). 

In DSA theory, a small fraction of incoming thermal particles can be injected into the CR population, 
and accelerated to very high energies through their interactions with resonantly scattering Alfv\'en waves 
in the converging flows across the SN shock (\eg Drury \etal 2001). 
Hence the strength of the turbulent magnetic field is one of the most important ingredients, 
which govern the acceleration rate and in turn the maximum energy of the accelerated
particles.
If the magnetic field strength upstream of SNRs is similar to the mean interstellar medium 
(ISM) field of $B_{\rm ISM} \sim 5 \mu$G, the maximum energy of CR ions of charge $Z$ is
estimated to be $E_{\rm max} \sim 10^{14}Z$ eV (Lagage \& Cesarsky 1983).
However, high-resolution X-ray observations of several young SNRs exhibit very thin rims,
indicating the presence of magnetic fields as strong as a few $100 \mu$G downstream of the shock
(\eg Bamba \etal 2003, Parizot \etal 2006). 
Moreover, theoretical studies have shown that efficient magnetic field amplification 
via resonant and non-resonant wave-particle interactions is an integral part of DSA 
(Lucek \& Bell 2000, Bell 2004).
If there exist such amplified magnetic fields in the upstream region of SNRs, 
CR ions might gain energies up to $E_{\rm max} \sim 10^{15.5}Z$ eV, 
which may explain the all-particle CR spectrum up to the second knee 
at $\sim10^{17}$ eV with rigidity-dependent energy cutoffs.
A self-consistent treatment of the magnetic field amplification 
has been implemented in several previous studies of nonlinear DSA (\eg Amato \& and Blasi 2006,
Vladimirov et al. 2008).

In Kang 2006 (Paper I, hereafter),
we calculated the CR acceleration at typical remnants from Type Ia supernovae 
expanding into a uniform interstellar medium (ISM).
With the upstream magnetic fields of $B_0=30\mu$G amplified by the CR streaming instability,
it was shown that the particle energy can reach up to $10^{16}Z$ eV at young
SNRs of several thousand years old, which is
much higher than what Lagage \& Cesarsky predicted. 
But the CR injection and acceleration efficiencies are reduced somewhat 
due to faster Alfv\'en wave speed.
With the particle injection fraction $\sim 10^{-4}-10^{-3}$,
the DSA at SNRs is very efficient, so that up to 40-50 \% of the explosion energy 
can be transferred to the CR component.
We also found that, for the SNRs in the warm ISM ($T_0=10^4$K), 
the accelerated CR energy spectrum should exhibit a concave curvature 
with the power-law slope, $\alpha$ (where $N(E)\propto E^{-\alpha}$) flattening
from 2 to 1.6 at $E > 0.1$ TeV.
In fact, the concavity in the CR energy spectrum is characteristic of strong ($M>10$) 
CR modified shocks when the injection fraction is greater than $10^{-4}$.  
(\eg Malkov \& Drury 2001, Berezhko \& V\"olk 1997, Blasi \etal 2005)

Recently, Ave \etal (2009) have analyzed the spectrum of CR nuclei up to $\sim 10^{14}$ eV 
measured by TRACER instrument and found that the CR spectra at Earth can be
fitted by a single power law of $J(E) \propto E^{-2.67}$.
Assuming an energy-dependent propagation path length ($\Lambda \propto E^{-0.6}$), they
suggested that a soft source spectrum, $N(E)$ with $\alpha
\sim 2.3-2.4$ is preferred by the observed data.
However, the DSA predicts that $\alpha=2.0$ for strong shocks in the test-particle limit
and even smaller values for CR modified shocks in the efficient acceleration regime
as shown in Paper I. 
Thus in order to reconcile the DSA prediction with the TRACER data
the CR acceleration efficiency at typical SNRs should be minimal and
perhaps no more than 10 \% of the explosion energy transferred to CRs (\ie test-particle
limit).
Moreover, recent Fermi-LAT observations of Cas A, which is only 330 years old
and has just entered the Sedov phase, 
indicate that only about 2\% of the explosion energy has been transfered to CR electrons and protons,
and that the soft proton spectrum with $E^{-2.3}$ is preferred to fit the observed gamma-ray 
spectrum (Abdo \etal 2010).
According to Paper I, such inefficient acceleration is possible only for SNRs in the hot
phase of the ISM and for the injected particle fraction smaller than $10^{-4}$.
One way to soften the CR spectrum beyond the canonical test-particle slope ($\alpha > 2$)
is to include the Alfv\'enic drift in the precursor, which reduces the velocity jump
across the shock. Zirakashvili \& Ptuskin (2008) showed that the Alfv\'enic
drift in the amplified magnetic fields both upstream and downstream can
drastically soften the accelerated particle spectrum.
We will explore this issue using our numerical simulations below. 

Caprioli \etal (2009) took a different approach to reconcile the concave
CR spectrum predicted by nonlinear DSA theory with the softer spectrum inferred from
observed $J(E)$.
They suggested that the CR spectrum at Earth is the sum of the time integrated flux 
of the particles that escape from upstream during the ST stage
and the flux of particles confined in the remnant and escaping at later times.
They considered several cases and found the injected spectrum could be softer
than the concave instantaneous spectrum at the shock.
The main uncertainties in their calculations are related with specific
recipes for the particle escape. It is not well understood at the present time how
the particles escape through a free escape boundary ($x_{\rm esc}$)
located at a certain distance upstream of the shock or through a maximum momentum 
boundary due to lack of (self-generated) resonant scatterings above an escape momentum.
The escape or release of CRs accelerated in SNRs to the ISM remains largely unknown
and needs to be investigated further.   

One of the key aspects of the DSA model is the injection process through which suprathermal
particles in the Maxwellian tail get accelerated and injected into the Fermi process.
However, the CR injection and consequently the acceleration efficiency still remain uncertain,
because complex interplay among CRs, waves, and the underlying gas flow
(\ie self-excitation of waves, resonant scatterings of particles by waves, and non-linear 
feedback to the gas flow) is all model-dependent and not understood completely. 

In this paper, we adopted two different injection recipes based on thermal leakage process,
which were considered previously by us and others. 
Then we have explored the CR acceleration at SNR shocks in the different temperature phases
(\ie different shock Mach numbers) and with different injection rates. 
Details of the numerical simulations and model parameters are described in \S II.
The simulation results are presented and discussed in \S III,
followed by a summary in \S IV.

\section{NUMERICAL METHOD}


\subsection{Spherical CRASH code}

Here we consider the CR acceleration at a quasi-parallel shock 
where the magnetic field lines are parallel to the shock normal. 
So we solve the standard gasdynamic equations with CR pressure terms
added in the Eulerian formulation for one dimensional spherical 
symmetric geometry.
The basic gasdynamic equations and details of the spherical CRASH (Cosmic-Ray Amr SHock) code 
can be found in Paper I and Kang \& Jones (2006).

In the kinetic equation approach to numerical study of DSA, 
the following diffusion-convection equation for the particle momentum 
distribution, $f(p)$, is solved along with suitably modified gasdynamic 
equations (\eg Kang \& Jones 2006):
\begin{eqnarray}
{\partial g\over \partial t}  + (u+u_w) {\partial g \over \partial r}
= {1\over{3r^2}} {\partial \over \partial r} \left[r^2 (u+u_w)\right]
\left( {\partial g\over \partial y} -4g \right) \nonumber\\
+ {1 \over r^2}{\partial \over \partial r} \left[r^2 \kappa(r,y)
{\partial g \over \partial r}\right],
\label{diffcon}
\end{eqnarray}
where $g=p^4f$, with $f(p,r,t)$ the pitch angle averaged CR
distribution,
and $y=\ln(p)$, and $\kappa(r,y)$ is the diffusion coefficient
parallel to the field lines \cite{skill75}.
So the proton number density is given by $n_{CR,p}= 4\pi \int f(p)p^2 dp$.
For simplicity we express the particle momentum, $p$ in units of $m_{\rm p}c$ 
and consider only the proton component.

The velocity $u_w$ represents the effective relative motion of
scattering centers with respect to the bulk flow velocity, $u$.
The mean wave speed is set to the Alfv\'en speed, \ie 
$u_w = v_A = B/ \sqrt{4\pi \rho}$ in the upstream region.
This term reflects the fact that
the scattering by Alfv\'en waves tends to isotropize
the CR distribution in the wave frame rather than the gas frame.
In the postshock region, $u_w = 0$ is assumed, since the Alfv\'enic
turbulence in that region is probably relatively balanced.
This reduces the
velocity difference between upstream and downstream scattering centers
compared to the bulk flow, leading to less efficient DSA.
This in turn affects the CR spectrum, and so
the `modified' test-particle slope can be estimated as 
\begin{equation}
q_{\rm tp} = {{3(u_0 - v_A)} \over {u_0-v_A -u_2}} 
\end{equation}
where $f(p)\propto p^{-q_{\rm tp}}$ is assumed (\eg Kang \etal 2009).
Hereafter we use the subscripts '0', '1', and '2' to denote
conditions far upstream of the shock, immediately upstream of the
gas subshock and immediately downstream of the subshock, respectively.
Thus the drift of Alfv\'en waves in the upstream region tends to soften
the CR spectrum from the canonical test-particle spectrum of $f(p)\propto p^{-4}$
if the Alfv\'en Mach number ($M_A= u_s/v_A$) is small.
We note $\alpha = q - 2$ for relativistic energies.
For example, for a strong shock with $u_2=u_0/4$ in the test particle limit, 
we can obtain the observed value of $\alpha=2.3$ if $v_A=0.173 u_0$.
 
Gas heating due to Alfv\'en wave dissipation in the upstream region is
represented by the term
\begin{equation}
W(r,t)= -  \omega_H v_A {\partial P_c \over \partial r },
\end{equation}
where 
$P_c= (4 \pi m_pc^2/3) \int g(p) dp/\sqrt{p^2+1}$ is the CR pressure.
This term is derived from a simple model in which Alfv\'en waves are amplified by
streaming CRs and dissipated locally as heat in the precursor region
(\eg Jones 1993).
As was previously shown in SNR simulations (\eg Berezhko \& V\"olk 1997,
Kang \& Jones 2006), precursor heating by wave dissipation reduces
the subshock Mach number thereby reducing DSA efficiency.
The parameter $\omega_H$ is introduced to control the degree of wave dissipation.
We set $\omega_H=1$ for all models unless stated otherwise.

\begin{table*}
\begin{center}
{\bf Table 1.}~~Model Parameters\\
\vskip 0.3cm
\begin{tabular}{ lrrrrrrrr }
\hline\hline
Model$^{\rm a}$ & $n_H$ (ISM) & $T_0$ & $E_0$ & ~$B_{\mu}$  & $r_o$ &~~$t_o$ & 
  $u_o$ &$P_o$  \\
~ & $(cm^{-3})$ & (K) & ($10^{51}$ ergs) & ($\mu$G)  & (pc) & (years) & 
($10^4 \kms$) & ($10^{-6}$erg cm$^{-3}$) \\

\hline
WA/WB  & 0.3   & $3.3\times 10^4$ & 1. & 30 & 3.19 & 255. &1.22 & 1.05 \\
MA/MB  & 0.03  &$10^5$            & 1. & 30 & 6.87 & 549. &1.22 & $1.05\times10^{-1}$ \\
HA/HB  & 0.003 &$10^6$            & 1. & 30 & 14.8 & 1182. &1.22 & $1.05\times10^{-2}$ \\

\hline
\end{tabular}
\end{center}
$^{\rm a}$ `W', `M', and 'H' stands for the warm, intermediate, and hot phase of the ISM,
respectively, while 
`A' and `B' stands for the injection recipes A and B, respectively, described in II (b).
\end{table*}

\begin{figure*}[t]
\vskip -0.5cm
\centerline{\epsfysize=14cm\epsfbox{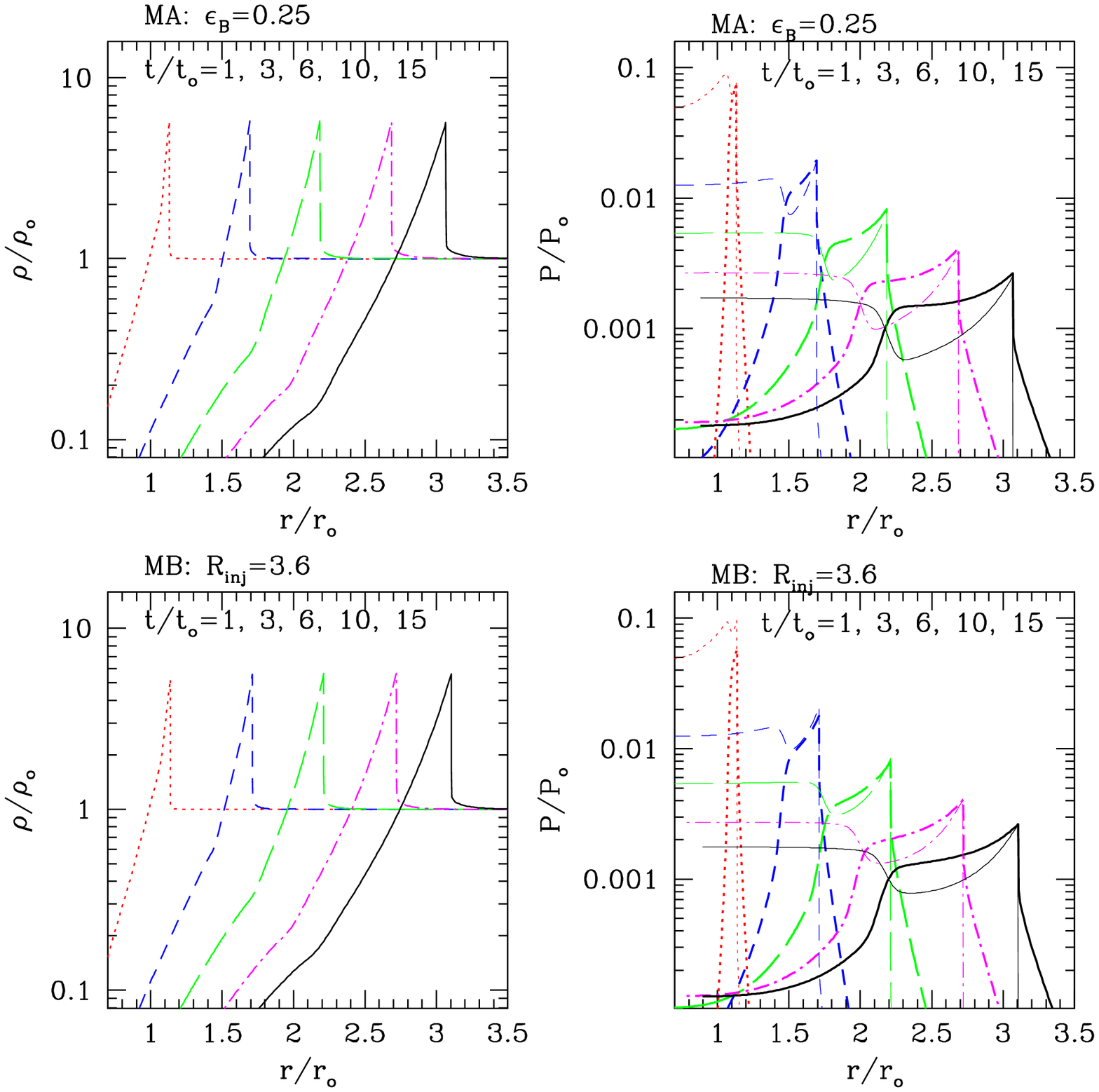}}
\vskip -0.5cm
\caption{
Time evolution of SNR model MA with $\epsilon_B=0.25$ (upper panels)
and SNR model MB with $R_{\rm inj}=3.6$ (lower panels)
at $t/t_o =  ~1.,~3.,~6.,~10.$ and 15.
In the right panels, heavy lines are for the CR pressure,
while thin lines are for the gas pressure.
The model parameters are
$M_{ej}=1.4 M_{\sun}$, $E_o=10^{51}$ ergs, $n_H=0.03 {\rm cm}^{-3}$,
$T_0=10^5$K, and $B_0=30\mu$G.
See Table 1 for the normalization constants.
}
\end{figure*}

Accurate solutions to the CR diffusion-convection equation
require a computational grid spacing significantly smaller 
than the particle diffusion length, $\Delta x \ll x_d(p) = \kappa(p)/u_s$.
With Bohm-like diffusion coefficient, $\kappa (p) \propto p$, 
a wide range of length scales must be resolved 
in order to follow the CR acceleration from the injection energy 
(typically $p_{\rm inj} \sim 10^{-2}$) to highly relativistic 
energy ($p \gg 1$). This constitutes an extremely challenging
numerical task, requiring rather extensive computational resources.
In order to overcome this difficulty, we have developed
CRASH code in 1D plane-parallel geometry 
(Kang \etal 2001) and in 1D spherical symmetric geometry (Kang \& Jones 2006)
by combining Adaptive Mesh Refinement technique and subgrid shock tracking 
technique.
Moreover, we solve the fluid and diffusion-convection 
equations in a frame comoving with the outer spherical shock
in order to implement the shock tracking technique effectively
in an expanding spherical geometry. 
In the comoving grid, the shock remains at the same location,
so the compression rate is applied consistently to the CR distribution
at the subshock, resulting in much more accurate and efficient 
low energy CR acceleration.

\subsection{Injection Recipes for Thermal Leakage}

The injection rate with which suprathermal particles are injected into CRs
at the subshock depends in general upon the shock Mach number, 
field obliquity angle, and strength of Alfv\'en turbulence responsible 
for scattering.
In thermal leakage injection models 
suprathermal particles well into the exponential tail of the postshock Maxwellian
distribution leak upstream across a quasi-parallel shock 
(Malkov \& V\"olk 1998; Malkov \& Drury 2001).
Currently, however, these microphysics issues are known poorly and any quantitative
predictions of macrophysical injection rate require extensive understandings
of complex plasma interactions. 
Thus this process has been handled numerically by adopting some 
phenomenological injection schemes in which the particles above a
certain injection momentum $p_{\rm inj}$ cross the shock and get
injected to the CR population. 

There exist two types of such injection models considered previously
by several authors.
In a simpler form, $p_{\rm inj}$ represents the momentum boundary
between thermal and CR population and so 
the particles are injected at this momentum 
(\eg Kang \& Jones 1995, Berezhko \& V\"olk 1997, Blasi \etal 2005).
The injection momentum is then expressed as
\begin{equation}
p_{\rm inj} = R_{\rm inj} p_{th},
\end{equation}  
where $R_{\rm inj}$ is a constant and $p_{th}= \sqrt{2k_B T_2 m_p}$ is
the thermal peak momentum of the Maxwellian distribution
of the immediate postshock gas with temperature $T_2$, 
and $k_B$ is the Boltzmann constant.
The CR distribution at $p_{\rm inj}$ is then fixed by the Maxwellian
distribution, 
\begin{equation}
f(p_{\rm inj})= n_2 \left({m_p \over {2\pi k_B T_2}}\right)^{1.5} \exp(-R_{\rm inj}^2),
\end{equation}  
where $n_2$ is the postshock proton number density.
Thus the constant parameter $R_{\rm inj}$ controls the injection rate
in this model. Here we refer this as `injection recipe B' and
consider the cases of $R_{\rm inj}=3.6$ and 3.8.

In Kang \etal (2002), on the other hand, a smooth ``transparency function'', 
$\tau_{\rm esc}(\epsilon_B, \upsilon)$ is adopted,
rather than a step-like filter function of the injection recipe B.
This function expresses the probability of supra-thermal
particles at a given velocity, $\upsilon$, leaking upstream through the
postshock MHD waves. 
One free parameter controls this function;  
$\epsilon_B = B_0/B_{\perp}$, which is the inverse ratio of the amplitude
of the postshock MHD wave turbulence $B_{\perp}$ to the general magnetic field
aligned with the shock normal, $B_0$ (Malkov \& V\"olk 1998).
In this model, the leakage probability $\tau_{esc}>0$ above $p_1 \approx 
m_p u_2 (1+1.07/\epsilon_B) \propto p_{th}$,
and the ``effective'' injection momentum is a few times $p_1$.
So the injection momentum can be expressed as
\begin{equation}
p_{\rm inj} = Q_{\rm inj}(M_s, \epsilon_B) p_{th}.
\end{equation} 
Note that the ratio $Q_{\rm inj}$ is a function of the subshock Mach number, $M_s$, 
as well as the parameter $\epsilon_B$,
while the constant ratio $R_{\rm inj}$ is independent of $M_s$.
The value of $Q_{\rm inj}$ is larger (and so the injection rate is smaller) 
for weaker subshocks and for smaller $\epsilon_B$ (see Kang \etal 2002).
In an evolving CR modified shock, the subshock weakens as the precursor develops
due to nonlinear feedback of the CR pressure and so the injection rate decreases in time.
We refer this as `injection recipe A'
and consider $0.2\le \epsilon_B\le 0.3$ here.

In Paper I we only considered the gas with protons (\ie mean molecular weight $\mu = 1$), 
but here we assume fully ionized plasma with cosmic abundance ($\mu = 0.61$).
As a result, for given gas pressure and density, the temperature
is lower and so slightly larger $\epsilon_B$ is needed to obtain
the similar level of injection as in Paper I.
Note that $\epsilon_B=0.16-0.2$ in Paper I.

The efficiency of the particle injection is quantified by
the fraction of particles swept through the shock 
that have been injected into the CR distribution:
\begin{equation}
\xi(t)=\frac {\int 4\pi r^2{\rm d r} \int 4\pi f(p,r,t)p^2 {\rm d p}}
{ \int 4\pi r_s^2 n_0 u_s {\rm dt }},
\end{equation}
where $n_0$ is the
proton number density far upstream and $r_s$ is the shock radius. 
Recent observations of nonthermal radiation from several SNRs indicate
that the injection fraction is about $\xi \sim 10^{-4}$ 
(\eg Berezhko \etal 2009, Morlino \etal 2009).

In our simulations, initially there is no pre-existing CRs and so all CR 
particles are freshly injected at the shock.

\subsection{A Bohm-like Diffusion Model}

Self-excitation of Alfv\'en waves by the CR streaming instability 
in the upstream region is an integral part of the DSA 
(Bell 1978; Lucek \& Bell 2004). 
The particles are resonantly scattered by those waves, diffuse
across the shock, and get injected into the Fermi first-order process.
These complex interactions are represented by the diffusion 
coefficient, which is expressed
in terms of a mean scattering length, $\lambda$, and the particle
speed, $\upsilon$, as
$\kappa(x,p) =  \lambda \upsilon/3$.
The Bohm diffusion model is commonly used to represent a saturated
wave spectrum (\ie $\lambda = r_g$, where $r_g$ is the gyro-radius), 
$\kappa_B(p) =  \kappa_n p^2/\ (p^2+1)^{1/2}$.
Here 
$\kappa_n= m c^3/(3eB)= 3.13\times 10^{22} {\rm cm^2s^{-1}} B_{\mu}^{-1}$,
and $B_{\mu}$ is the magnetic field strength in units of microgauss.
As in Paper I, we adopt a Bohm-like diffusion coefficient
that includes a weaker non-relativistic momentum dependence,
\begin{equation}
\kappa(r,p) = \kappa_{\rm n}\cdot p {\rho_0 \over \rho(r)}.
\end{equation}
Since we do not follow explicitly the amplification of magnetic fields due to
streaming CRs,
we simply assume that the field strength scales with compression and
so the diffusion coefficient scales inversely with density.

\section{Simulations of Sedov-Taylor Blast Waves}
\begin{figure*}[t]
\vskip -0.5cm
\centerline{\epsfysize=14cm\epsfbox{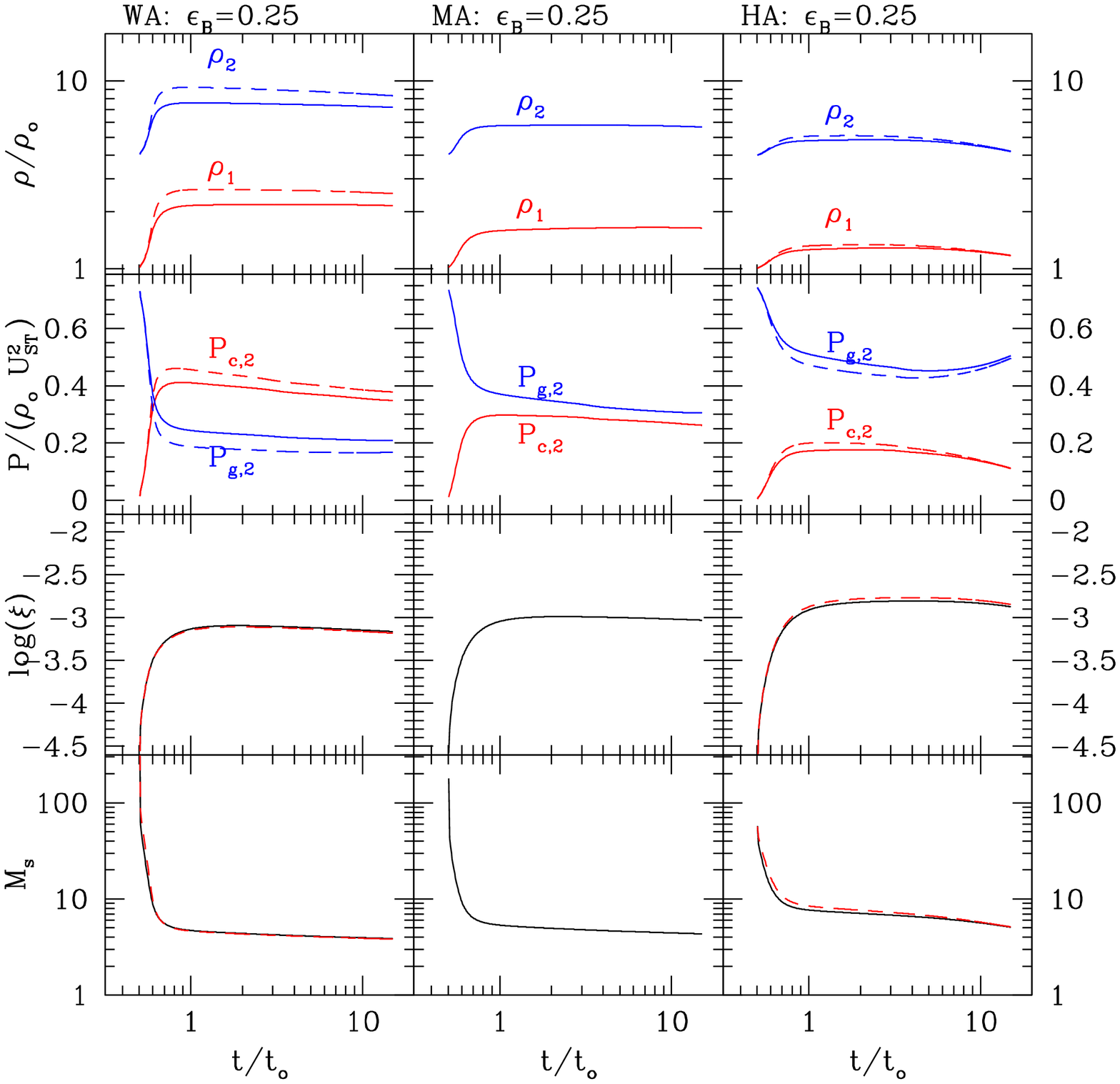}}
\vskip -0.5cm
\caption{
Immediate pre-subshock density, $\rho_1$, post-subshock density, $\rho_2$, 
post-subshock CR and gas pressure in units of the ram pressure of the 
unmodified Sedov-Taylor solution, $\rho_0 U_{ST}^2 \propto (t/t_o)^{-6/5}$, 
the CR injection parameter, $\xi$, and subshock Mach number, $M_s$ 
are plotted for models WA (left panels), MA (middle panels), and 
HA (right panels).
See Table 1 for the model parameters.
The injection recipe A with $\epsilon_B=0.25$ is adopted.
For WA and HA models, the dashed lines are for the runs with
a reduced wave heating parameter, $w_H=0.5$.
}
\end{figure*}
\begin{figure*}[t]
\vskip -0.5cm
\centerline{\epsfysize=14cm\epsfbox{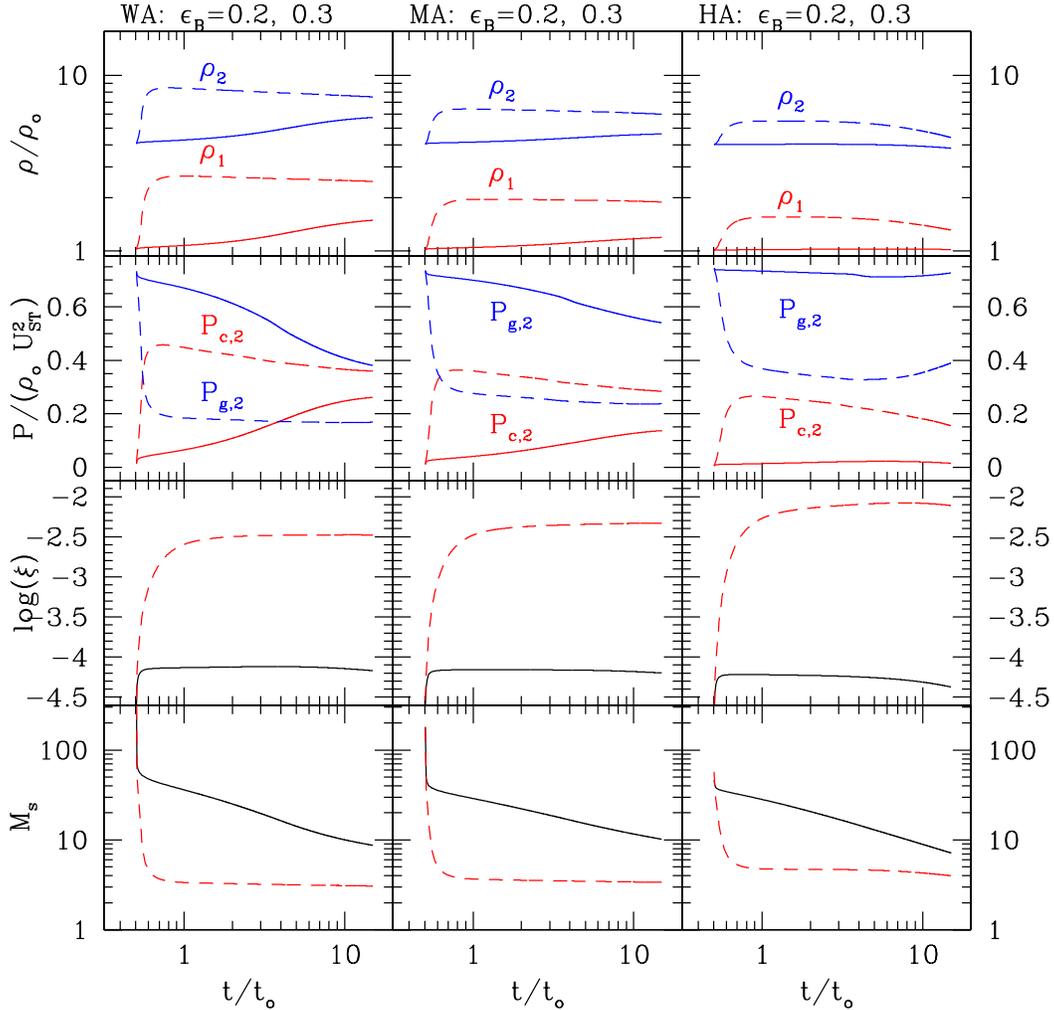}}
\vskip -0.5cm
\caption{
The same as Fig. 2 except that $\epsilon_B=0.2$ (solid lines) or 
0.3 (dashed lines) is adopted
}
\end{figure*}
\begin{figure*}[t]
\vskip -0.5cm
\centerline{\epsfysize=14cm\epsfbox{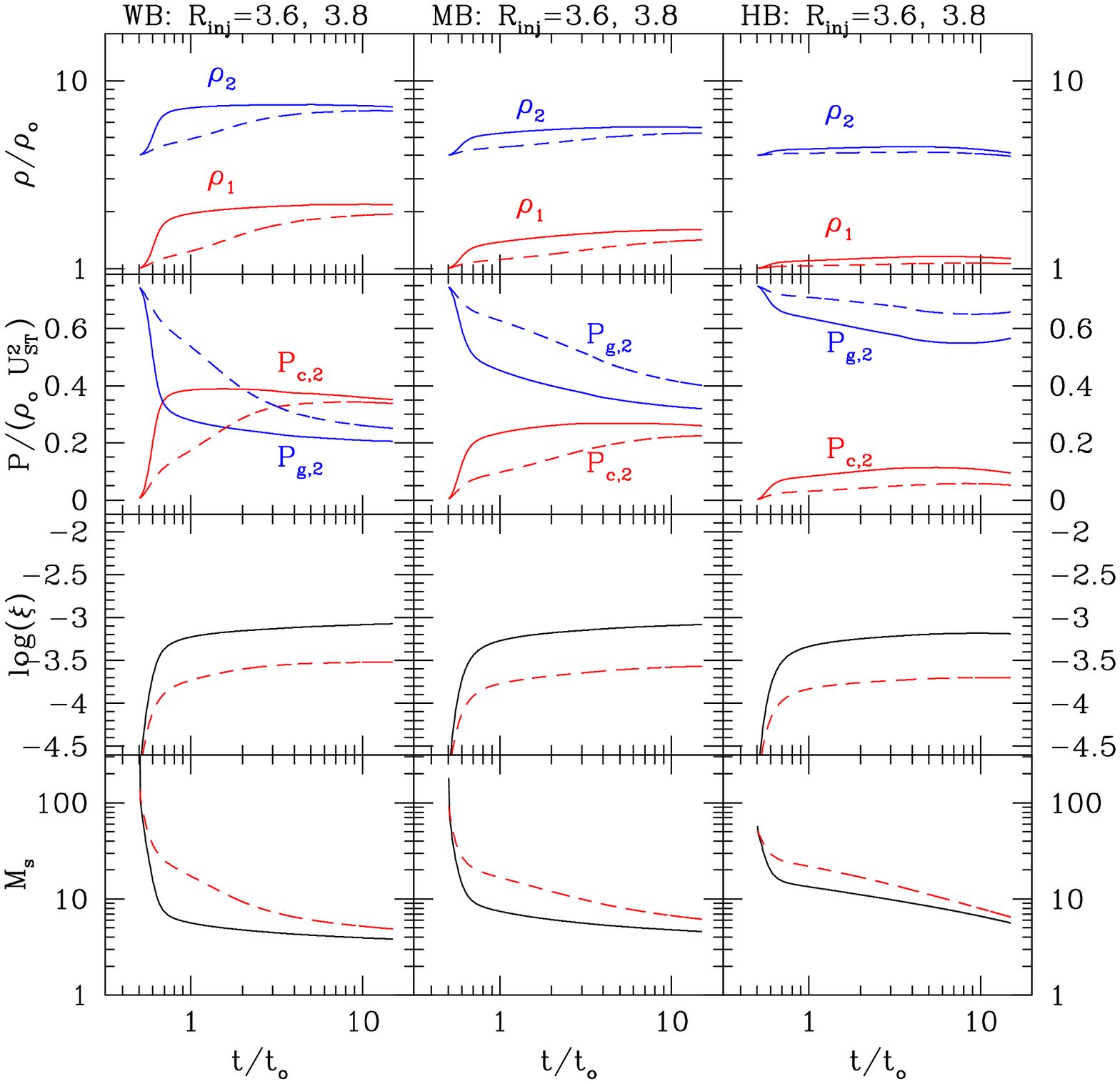}}
\vskip -0.5cm
\caption{
The same as Fig. 2 except that the injection recipe B with
$R_{\rm inj}=3.6$ or $R_{\rm inj}=3.8$ is adopted
}
\end{figure*}

\begin{figure*}[t]
\vskip -0.5cm
\centerline{\epsfysize=14cm\epsfbox{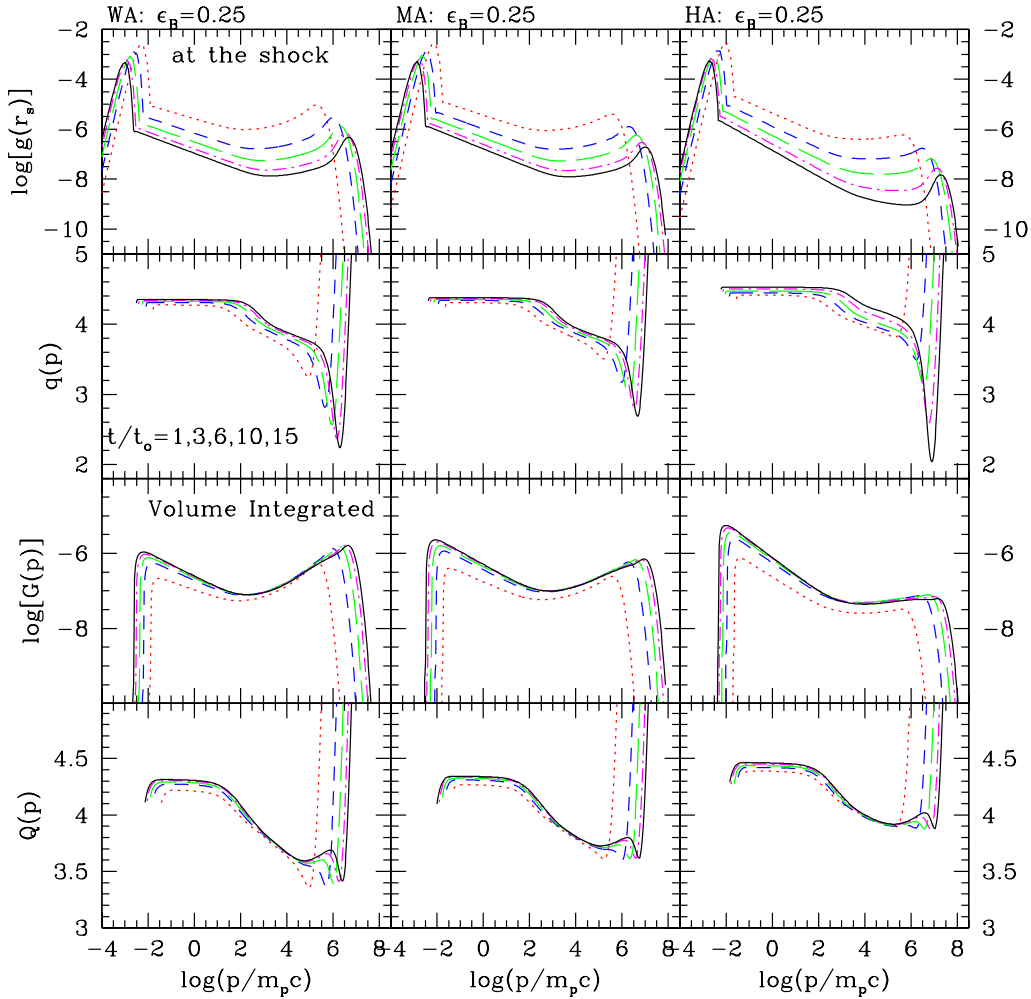}}
\vskip -0.5cm
\caption{
The CR distribution at the shock, $g(r_s,p)$, and
its slope, $q(p) = - d (\ln g(r_s,p))/ d \ln p + 4 $, 
the volume integrated CR number, 
$G(p)= \int g(r,p) 4\pi r^2 {\rm d}r$, and
its slope, $Q(p) = - d (\ln G(p))/ d \ln p + 4 $, 
are shown at $t/t_o=$ 1, 3, 6, 10, and 15
for models WA (left panels), MA (middle panels), and 
HA (right panels).
The injection recipe A with $\epsilon_B=0.25$ is adopted.
}
\end{figure*}
\begin{figure*}[t]
\vskip -0.5cm
\centerline{\epsfysize=14cm\epsfbox{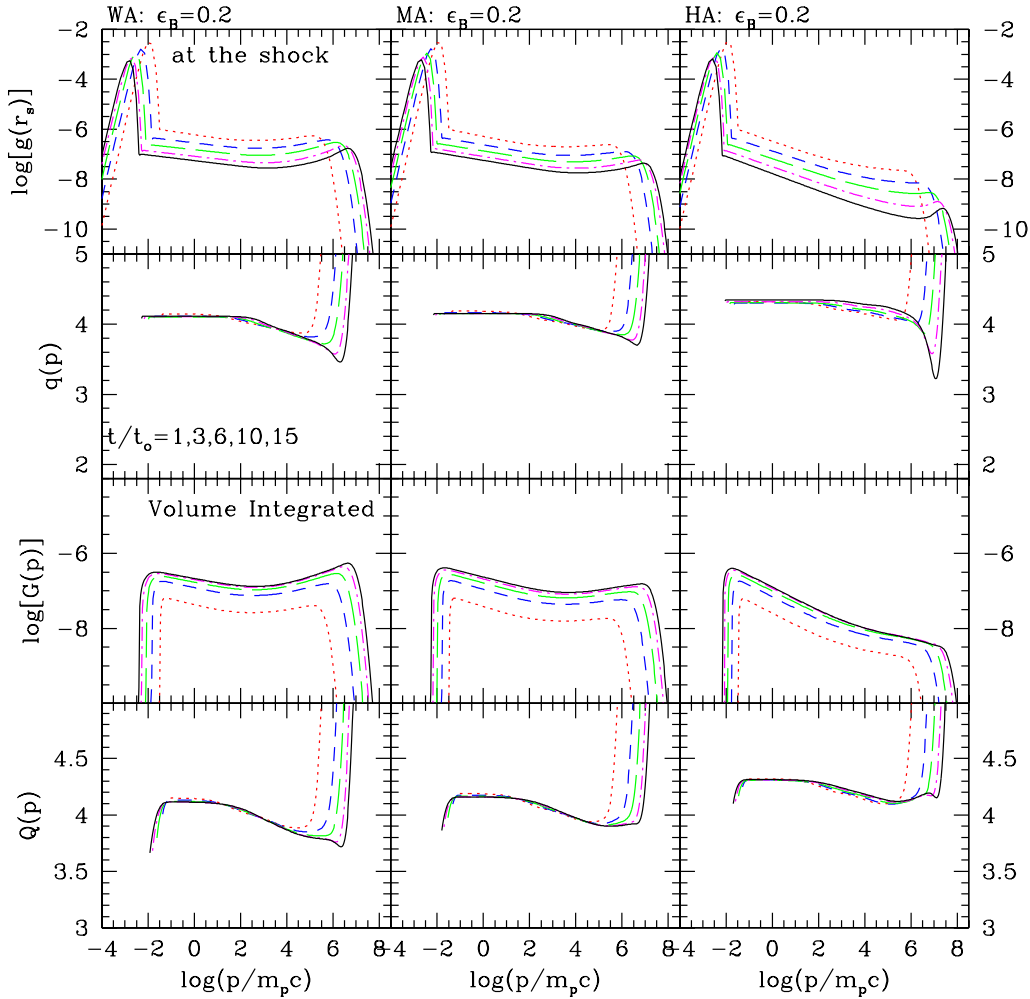}}
\vskip -0.5cm
\caption{
The same as Fig. 5 except that $\epsilon_B=0.2$ is adopted.
The injection and acceleration efficiencies are low. 
}
\end{figure*}
\begin{figure*}[t]
\vskip -0.5cm
\centerline{\epsfysize=14cm\epsfbox{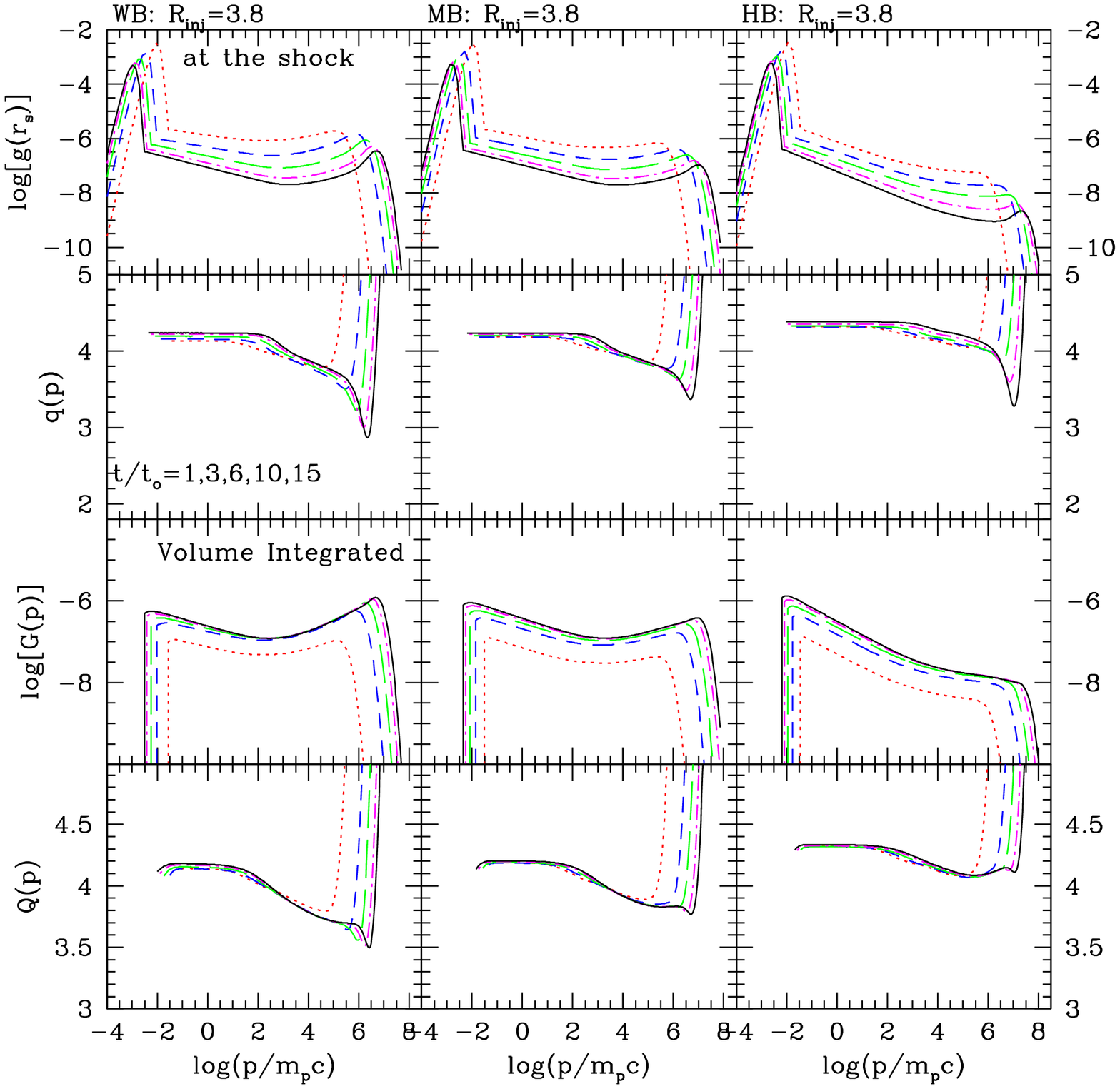}}
\vskip -0.5cm
\caption{
The same as Fig. 5 except that the injection recipe B with $R_{\rm inj}=3.8$ 
is adopted.
}
\end{figure*}

\subsection{SNR Model Parameters}

As in Paper I, we consider a Type Ia supernova explosion with the ejecta mass, 
$M_{ej}=1.4 M_{\sun}$, expanding into a uniform ISM.
All models have the explosion energy, $E_o=10^{51}$ ergs. 
Previous studies have shown that 
the shock Mach number is the key parameter determining the evolution and the DSA efficiency, 
although other processes such as the particle injection, the Alfv\'enic drift and dissipation 
do play certain roles (e.g. Kang \& Jones 2002, 2007).
So here three phases of the ISM are considered:
the {\it warm phase} with $n_H=0.3{\rm cm}^{-3}$ and $T_0=3\times 10^4$K,
the {\it hot phase} with $n_H=0.003~ {\rm cm}^{-3}$ and $T_0=10^6$K,
and the {\it intermediate phase} with
$n_H=0.03~ {\rm cm}^{-3}$ and $T_0=10^5$K.
The pressure of the background gas is
$P_{\rm ISM} \sim 10^{-12}~{\rm erg ~cm^{-3}}$.
Model parameters are summarized in Table 1.
For example, `WA' model stands for the warm phase and the injection recipe A,
while `MB' model stands for the intermediate phase and the
injection recipe B. 

Recent X-ray observations of young SNRs indicate a magnetic field
strength much greater than the mean ISM field of $5\mu$G
(\eg Berezhko \etal 2003; V\"olk \etal 2005).
Thus, to represent this effect we take the upstream field strength, $B_0=30\mu$G.
The strength of magnetic field determines the size of
diffusion coefficient, $\kappa_n$, and the drift speed of Alfv\'en
waves relative to the bulk flow.
The Alfv\'en speed is given by $v_A= v_{A,0}(\rho/\rho_0)^{-1/2}$
where $v_{A,0} = (1.8 {~\rm km s^{-1}})B_{\mu}/\sqrt{n_H}$.
So in the hot phase of the ISM (HA/HB models), $v_{A,0}=986 {~\rm km s^{-1}}$
and $v_{A,0}/u_s \approx 0.175$ at $t=t_o$, 

The physical quantities are normalized, both in the numerical code and in
the plots below, by the following constants:
\begin{eqnarray}
r_o=\left({3M_{ej} \over 4\pi\rho_o}\right)^{1/3},\nonumber\\
t_o=\left({\rho_o r_o^5 \over E_o}\right)^{1/2}, \nonumber\\
u_o=r_o/t_o,\nonumber\\ 
\rho_o = (2.34\times 10^{-24} {\rm g cm^{-3}}) n_H,\nonumber\\ 
P_o=\rho_o u_o^2.\nonumber
\end{eqnarray}
These values are also given in Table 1 for reference.
Note that these physical scales depend only on $n_H$, since
$M_{ej}$ and $E_o$ are the same for all models.

For a SNR propagating into a uniform ISM, 
the highest momentum, $p_{\rm max}$, is achieved at the beginning of the 
Sedov-Taylor (ST hereafter) stage and the transfer of explosion energy to the CR 
component occurs mostly during the early ST stage (\eg Berezhko \etal 1997).
In order to take account of the CR acceleration from free expansion
stage through ST stage, we begin the calculations with the 
ST similarity solution at $t/t_o=0.5$ and terminated them
at $t/t_o=15$. 
See Paper I for further discussion on this issue.

\begin{figure*}[t]
\vskip -0.5cm
\centerline{\epsfysize=14cm\epsfbox{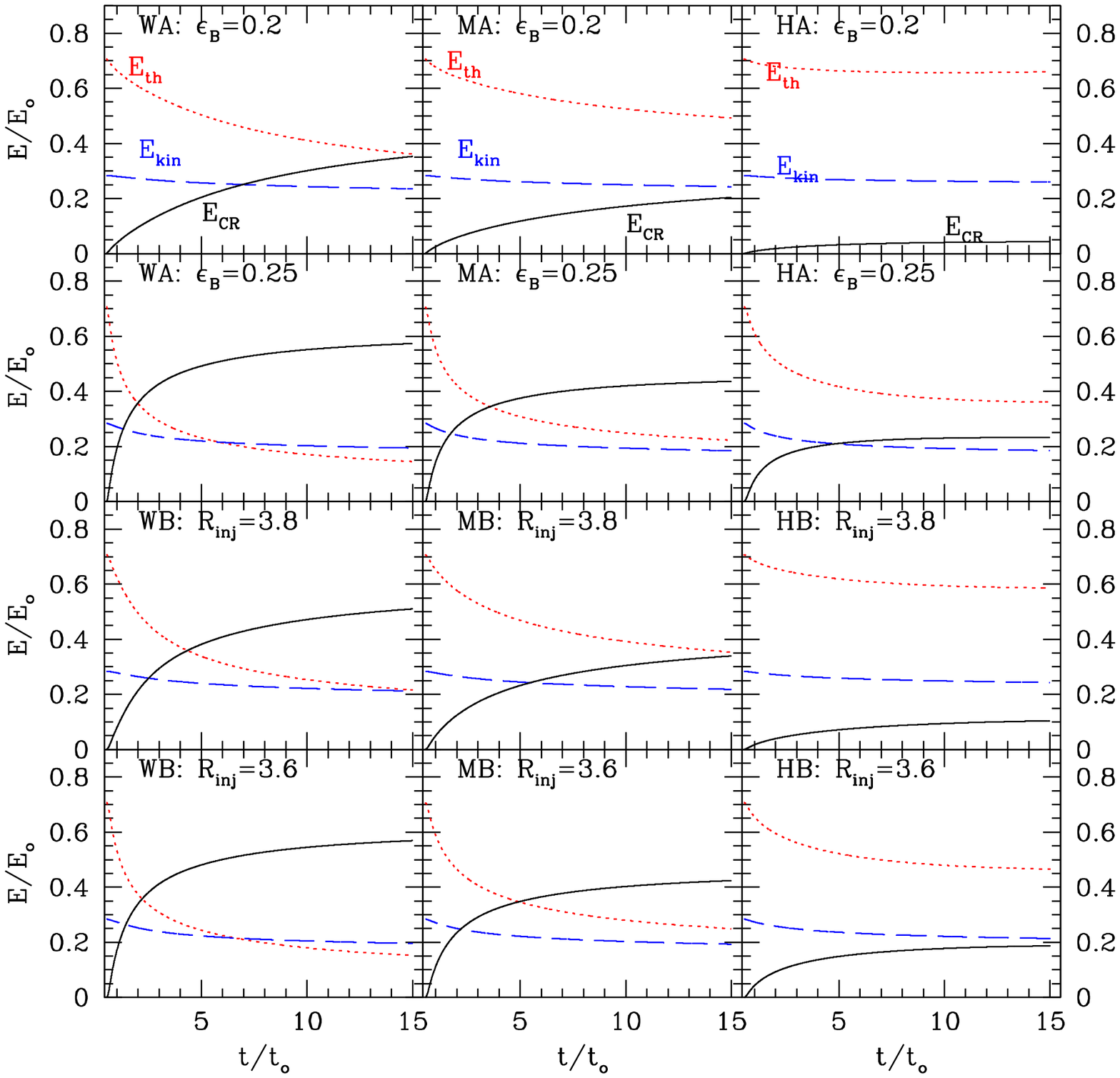}}
\vskip -0.5cm
\caption{
Integrated thermal, kinetic and CR energies inside the simulation volume
as a function of time for different models.
The injection parameter is from top to bottom,
$\epsilon_B=0.2,$ $\epsilon_B= 0.25$, $R_{\rm inj}=3.8$, and $R_{\rm inj}=3.6$.
See Table 1 for model parameters.
}
\end{figure*}

\section{RESULTS}
\subsection{Remnant Evolution}

Fig. 1 shows the evolution of SNRs in the intermediate temperature phase  
with $\epsilon_B=0.23$ (injection recipe A) and with $R_{\rm inj}=3.6$ 
(injection recipe B). 
The spatial profile and the evolution of SNRs are quite similar in the two
models, implying that detail difference between the two injection recipes 
is not crucial.
This is because the subshock Mach number reduces to $M_s\approx 4$ at the
early stage and remains the same until the end of simulations in the both
models, resulting in similar evolutionary behavior of $\xi$ (see Fig. 2). 
In these efficient acceleration models, by the early ST stage, $t/t_o \sim 1$, 
the forward shock has already become dominated by the CR pressure 
and the total density compression ratio becomes $\rho_2/\rho_0 \approx 6$ 
in both models.
Spatial distribution of the CR pressure widens and becomes broader than
that of the gas pressure at the later stage.

The precursor length scale is given by the diffusion length of the
highest momentum particles, $l_{\rm d,max} \approx 0.1 \int u_s(t) dt$,
independent of the diffusion coefficient $\kappa_n$ (Kang \etal 2009).
In the test-particle limit, the shock would follow
the ST similarity solution given by
\begin{equation} 
U_{\rm ST}/u_o= 0.4\xi_s (t/t_o)^{-0.6}, 
\end{equation} 
where $\xi_s=1.15167$ is the similarity parameter (Spritzer 1978).
Then $l_{\rm d,max}/r_o \approx 0.1 \xi_s(t/t_o)^{0.4}$.
Since the shock radius of the ST solution is $r_{\rm ST}/r_o = \xi_s(t/t_o)^{0.4}$,
the relative width of the precursor 
is estimated to be $l_{\rm d,max}/r_s \approx 0.1$,
independent of $\kappa_n$.
In Fig. 1 one can see narrow precursors 
in the density and CR pressure distribution,
consistent with this estimate.

The shock Mach number is the primary parameter that determines the
CR acceleration efficiency,
while the injection parameters $\epsilon_B$ and $R_{\rm inj}$ are the
secondary parameters.
So the temperature (\ie sound speed) of the background ISM is important.
The Mach numbers of the initial shock are $M_{s,i} \approx $ 310, 180, and 60
in the warm, intermediate, hot ISM models, respectively. 
For the warm (WA/WB) and intermediate (MA/MB) models, 
the CR acceleration is efficient with the postshock
CR pressure, $0.2\la P_{c,2}/(\rho_0 U_{ST}^2) \la 0.4$, and 
the shock is significantly modified.
We will refer these models as `efficient acceleration models'.
For HA/HB models, the CR acceleration is inefficient
with $P_{c,2}/(\rho_0 U_{ST}^2) < 0.1$ 
and the shock is almost test-particle like. 
So the hot ISM models are `inefficient acceleration models'.
Regarding the injected particle fractions,
the models with $\epsilon_B\gsim 0.23$ or $R_{\rm inj}=3.6-3.8$ have
the injection fraction, $\xi > 10^{-4}$ and represent `efficient injection models'.
The models with $\epsilon_B= 0.2$ have $\xi \approx 10^{-4.2}$ and 
are `inefficient injection models'.

Figs. 2-4 show the evolution of shock properties such as
the compression factors, postshock pressures, the injection fraction, 
and subshock Mach number for various models.
In Fig. 2, the models with $\epsilon_B=0.25$ are shown for $\omega_H=0.5$ (dashed
lines) and $\omega_H=1.0$ (solid lines).
We can see that the more efficient wave dissipation (\ie larger $\omega_H$)
reduces the CR acceleration and the flow compression.
Here WA and MA models are characterized by both efficient injection and 
efficient acceleration, while HA models show an efficient injection but inefficient acceleration.
Most of shock properties seem to approach to more or less 
time-asymptotic values before $t/t_o=1$.   
As the precursor grows, the subshock weakens to $3\le M_s \le 5$ 
in these models. 
The injected CR particle fraction is about $\xi \approx 10^{-3}$.
The postshock CR pressure is about $P_{c,2}/(\rho_0 U_{ST}^2) \approx 0.4$,
0.25, and 0.1 for WA, MA, and HA models, respectively.
The compression factor in the precursor varies a little among
different models, typically $\rho_1/\rho_0\approx 2-3$. 
The total density compression is
$\rho_2/\rho_0 \approx$ 7-8 for WA models, 5.5 for MB model,
4.5 for HA models, indicating the CR modified shock structure.


Comparison of Figs. 2 and 3 tells us how the CR acceleration depends on
the injection parameter $\epsilon_B$ and consequently on $\xi$.
The injection fraction is 
$\xi \approx 10^{-4.2}$ for $ \epsilon_B= 0.2$ (inefficient injection models), 
$\xi \approx 10^{-3} - 10^{-3.5}$ for $0.23\le \epsilon_B \le 0.25$, 
and $\xi \approx 10^{-2.5}-10^{-2}$ for $\epsilon_B=0.3$. 
In the inefficient injection models, the subshock Mach number and
postshock properties change rather gradually throughout the Sedov stage
and the shock is almost test-particle like with $\rho_1/\rho_0\approx 1$.
On the other hand, for $\epsilon_B=0.3$ the injection fraction seems much higher than 
what is inferred from recent observations of nonthermal emission from young SNRs
(\eg Morlino \etal 2009).
The postshock CR pressure, $P_{c,2}/(\rho_0 U_{ST}^2)$, is
roughly independent of the injection fraction as long as $\xi \gsim 10^{-3.5}$
($\epsilon_B\gsim 0.23$).
But, for inefficient injection models with $\epsilon_B=0.2$, this ratio is reduced significantly.

We can see that the models with $\epsilon_B=0.25$ (Fig. 2) and 
the models with $R_{\rm inj}=3.6$ (Fig. 4) have similar results. 
This confirms that the shock Mach number is the primary factor that
controls the CR acceleration.
For the injection recipe B, the injection rate does not 
depend on the subshock Mach number, so the evolution of $\xi$ is similar among WB, MB,
and HB models.
The models with $R_{\rm inj}=3.8$ have about 3 times smaller injection fraction 
and evolve more slowly, compared to those with $R_{\rm inj}=3.6$.
But both models seems to approach to the similar states at $t/t_o>10$.

\subsection{Cosmic Ray Spectrum}

The rate of momentum gain for a particle under going DSA is given by
\begin{equation}
\frac{dp}{dt} = \frac{u_0-u_2}{3} \left ( \frac{u_0}{\kappa_0} + \frac{u_2}{\kappa_2} \right ) p.
\label{dpacc}
\end{equation}
Assuming that the shock approximately follows the ST solution 
and that the compression ratio is about four,
the maximum momentum of protons accelerated from $t_i$ to $t$,
can be calculated as 
\begin{equation}
p_{\rm max} \approx \frac{0.53 u_o^2 t_o}{ \kappa_n} \left [ \left ( \frac{t_i}{t_o} \right )^{-0.2} - \left ( \frac{t}{t_o} \right )^{-0.2} \right ].
\label{pmax}
\end{equation}
For our simulations started from $t_i/t_o = 0.5$,
this asymptotes to $p_{\rm max} \approx 0.61(u_o^2 t_o/\kappa_n) \sim 10^{6.5}$
at large $t$, which corresponds to $E_{\rm max} \approx 10^{15.5}$ eV.

Figs. 5-7 show the CR distribution function at the shock, $g(r_s,p)$,
and its slope  $q(p) = - d (\ln g(r_s,p))/ d \ln p +4 $,
and the volume integrated CR spectrum,
$G(p) = \int 4\pi g(r,p) r^2 {\rm d}r$ 
and its slope
$Q(q) = - d (\ln G(p))/ d \ln p + 4 $. 
The thermal population is included in the plot of $g(r_s,p)$ in order to demonstrate
how it is smoothly connected with the CR component through thermal leakage.
For the volume integrated spectrum, only the CR component is shown. 
We note that in our simulations the particles escape from the simulation 
box only by diffusing upstream and no escape condition is enforced.    
Thus $G(p)$ represents the spectrum of the particles confined by the shock, 
including the particles in the upstream region.
From the spectra in Figs. 5-7, we can see that
$p_{\rm max}$ approaches up to $\sim 10^{15}-10^{15.5}$ eV/c 
at $t/t_o=15$ for all models, which is consistent with the estimate
given in Eq. (11). 

In Fig. 5, 
the CR spectra at the shock in the models with $\epsilon_B=0.25$
(high injection rate with $\xi\approx 10^{-3}$) 
exhibit the canonical nonlinear concave curvature.
This is a consequence of the following two effects: 
the large compression factor across the shock structure 
and the decreasing injection rate due to the slowing of the shock speed.
With the CR modified flow structure, the slope near $p_{\rm max}$
becomes harder with $q_t= 3 s \sigma_t/(s \sigma_t-1)$,
where $s=1-\upsilon_A/u_s$ is the modification factor due to the
Alfv\'enic drift and $\sigma_t= \rho_2/\rho_0 \gg 4$ 
is the total shock compression ratio.
If the subshock Mach number reduces to $M_s\approx 3-5$, 
the test-particle slope at low momenta becomes 
$q_s = 3 s \sigma_s/(s \sigma_s-1)\approx 4.2-4.5$, 
where $\sigma_s=\rho_2/\rho_1$ is the compression ratio across the subshock.
The particle flux through the shock, $\rho_0 u_s$, decreases, 
because the SNR shock slows down in time.
At the same time the injection rate decreases 
because the injection process is less efficient at weaker shocks. 
The combined effects result in the reduction of the amplitude of $f(r_s,p)$ near $p_{\rm inj}$.
Consequently, the CR spectrum at lower momentum steepens and decreases  in time.
Fig. 5 demonstrates that the modified flow structure along with slowing down of the shock speed 
accentuates the concavity of the CR spectrum 
in much higher degrees than what is normally observed in plane-parallel shocks.

However, the volume integrated spectrum $G(p)$ is more relevant 
for unresolved observations of SNRs or for the total CR spectrum produced by SNRs.
The concavity of $G(p)$ is much less pronounced than that 
of $g(r_s,p)$, and its slope $Q(p)$ varies 4.2-4.4
at low momentum and 3.5-4.0 at high momentum among different models. 
We see that $G(p)$ stays almost constant for $t/t_o \gsim 6$, 
especially for $10^{11}<E<10^{15}$eV, while extending in the momentum space
with decreasing $p_{\rm min}$ and increasing $p_{\rm max}$. 
This can be understood as follows.
From Figs. 2-4, $P_{c,2}/(\rho_0 U_{\rm ST}^2)\sim $constant for $t/t_o>1$
(except in the inefficient injection model with $\epsilon_B=0.2$), so
the CR pressure evolves like $P_{c,2} \propto t^{-6/5}$ (see also Fig. 1). 
The total CR energy associated with the remnant
is roughly $E_{\rm CR} \propto P_c R_{\rm ST}^3 \sim$ constant.
Since $E_{\rm CR} \propto \int G(p)dp$ approximately, so $G(p)$ should approach
to an asymptotic form for $t/t_o \gg 1$.
In other words, the distribution function $g(r/r_s,p)$ decreases as $t^{-6/5}$
in terms of the normalized coordinate, $r/r_s$, but the volume occupied by
the remnant increases as $t^{6/5}$, resulting in more or less constant $G(p)$.
Using this and the fact that $p_{\rm max}$ asymptotes to $0.6(u_o^2 t_o/\kappa_n)$ at
large $t$, we can predict that the form of $G(p)$ would remain about the
same at much later time.

As discussed in the Introduction,
the CR spectrum observed at Earth is $J(E) \propto E^{-2.67}$ for $10^9 < E < 10^{14}$ eV. 
This implies that the source spectrum should be roughly $N(E) \propto E^{-\alpha}$
with $\alpha = 2.3-2.4$,
if we assume an energy-dependent path length, $\Lambda(R) \propto R^{-0.6}$ 
(where $R=pc/Ze$ is the rigidity) (Ave \etal 2009).
If in fact the CR source spectrum at SNRs, $G(p)$, is assumed to be 
released into the ISM at the end of ST stage,
$N(E)dE \propto G(p)p^2dp$ is too flat to be consistent with the observation.
Thus from the spectra $G(p)$ in Fig. 5
we can infer that SNRs expanding into warm or intermediate phases of the 
ISM cannot be the dominant sources of Galactic CRs.

Even with the hot ISM models, the canonical test-particle spectrum,
$N(E)\propto E^{-2}$ might be still too flat.
If we consider the effects of Alfv\'en wave drift, however, the modified
test-particle slope will be given by Eq. (2) for strong plane-parallel shocks.
One can estimate that $v_A \approx 1000 {\rm km s^{-1}}$ for $n_H=0.003 {\rm
cm^{-3}}$ and $B_0=30 \mu$G, which leads to $\alpha \approx 2.3$.  
We show in Fig. 6 the CR spectra for inefficient injection models with $\epsilon_B=0.2$.
The spectra are less flat, compared with those of efficient injection models shown in Fig. 5.
Especially, the HA model with $\epsilon_B=0.2$ has the slope, $\alpha = 2.1-2.3$
for $10^{11}< E< 10^{15}$ eV.
This could be more compatible with observed $J(E)$ at Earth.

Fig. 7 shows the spectra for the models with injection recipe B ($R_{\rm inj}=3.8$).
Again $G(p)$ of HB model shows the slope, $\alpha=2.1-2.3$, 
for $10^{11}< E< 10^{15}$ eV.
In fact these spectra are quite similar to those for HA model shown in Fig. 6.

\subsection{Energy Conversion Factor}
Finally, Fig. 8 shows the integrated energies, 
$E_i/E_o = 4\pi \int e_i r^2 {\rm d}r$, where $e_{th}$,
$e_{kin}$, and $e_{CR}$ are the densities of thermal, kinetic and
cosmic ray energy, respectively.
The kinetic energy reduces only slightly and is similar for all models.
The total CR energy accelerated by $t/t_o= 15$ is
$E_{\rm CR}/E_o=$ 0.35, 0.20, and 0.05 for WA, MA, and HA models, respectively,
for $\epsilon_B=0.2$.
In the efficient injection models with $\epsilon_B=0.25$ or $R_{\rm inj}=3.6$,
the evolution of SNRs is quite similar, and the CR energy fraction approaches to
$E_{\rm CR}/E_o=$ 0.56, 0.43, and 0.25 for WA/WB, MA/MB, and HA/HB models, respectively. 
So in terms of the energy transfer fraction, the CR acceleration in the warm ISM models 
seems to be too efficient.
But one has to recall that the CR injection rate may depend 
on the mean magnetic field direction relative to the shock surface.
In a more realistic magnetic field geometry, where a uniform
ISM field is swept by the spherical shock, only 10-20 \% of the
shock surface has a quasi-parallel field geometry (V\"olk \etal 2003).
If the injection rate were to be reduced significantly at perpendicular shocks, 
one may argue that the CR energy conversion factor averaged over the entire
shock surface could be several times smaller than the factors shown in Fig. 8. 

On the other hand, Giacalone (2005) showed that the protons can
be injected efficiently even at perpendicular shocks in fully
turbulent fields due to field line meandering.
In such case the injection rate at perpendicular shocks may 
not be much smaller than that at parallel shocks
and the CR energy conversion may be similar.
Then SNRs in the warm phase of the ISM seem to generate too much CR energy. 
In order to meet the requirement of 10 \% energy conversion
and at the same time to reconcile with the CR spectrum observed at Earth,
SNRs expanding into the hot phase of the ISM should be the dominant 
accelerators of Galactic CRs below $10^{15}$eV.

\section{SUMMARY}

The evolution of cosmic ray modified shocks depends on complex
interactions between the particles, waves in the magnetic field, 
and underlying plasma flow.
We have developed numerical tools that can emulate some of those
interactions and incorporated them into a kinetic numerical 
scheme for DSA, CRASH code (Kang \etal 2002, Kang \& Jones 2006). 
Specifically, we assume that a Bohm-like diffusion arises due to
resonant scattering by Alfv\'en waves self-excited by the CR streaming 
instability, and adopt simple models for the drift and dissipation of
Alfv\'en waves in the precursor (Jones 1993; Kang \& Jones 2006).

In the present paper, using the spherical CRASH code, 
we have calculated the CR spectrum
accelerated at SNRs from Type Ia supernova expanding into a uniform
interstellar medium.
We considered different temperature phases of the ISM, since the shock
Mach number is the primary parameter that determines the
acceleration efficiency of DSA.
One of the secondary parameters is the fraction of particles 
injected into the CR population, $ \xi$, at the gas subshock.
Since detailed physical processes that governs the injection
are not known well, we considered two injection recipes that are
often adopted by previous authors.

The main difference between the two recipes is whether the ratio
of injection momentum to thermal peak momentum, \ie 
$p_{\rm inj}/p_{th}$, is constant or depends on the subshock Mach number.
It turns out the CR acceleration and the evolution of SNRs
are insensitive to such difference as long as the injection fraction is similar.
For example, the models with injection recipe A with $\epsilon_B=0.23$
and the models with injection recipe B with $R_{\rm inj}=3.6$ show almost
the same results with similar injection fractions, $\xi \approx 10^{-3.5}
-10^{-3}$.

In general the DSA is very efficient for strong SNR shocks, if
the injection fraction, $\xi \gsim 10^{-3.5}$.
The CR spectrum at the subshock shows a strong concavity, not only
because the shock structure is modified nonlinearly by the dominant
CR pressure, but also because the SNR shock slows down in time during the ST
stage.  Thus the concavity of the CR spectrum in SNRs is more 
pronounced than that in plane-parallel shocks.
However, the volume integrated spectrum, $G(p)$, (\ie the spectrum
of CRs confined by the shock including the particles in the upstream
region) is much less concave, which is consistent with previous
studies (\eg Berezhko \& V\"olk 1997).
We have shown also that $G(p)$ approaches roughly to time-asymptotic states,
since the CR pressure decreases as $t^{-6/5}$ while the volume increases
as $R_{\rm ST}^3 \propto t^{6/5}$.  
This in turn makes the total CR energy converted ($E_{\rm CR}$) asymptotes to a constant value. 
If we assume that CRs are released at the break-up of SNRs, then
the source spectrum can be modeled as $N(E)dE=G(p)p^2dp$.
However, it is a complex unknown problem how to relate $G(p)$ to the source
spectrum $N(E)$ and further to the observed spectrum $J(E)$. 
 
In the warm ISM models ($T_0=3\times10^4$K, $n_H=0.3{\rm cm^{-3}}$),
the CR acceleration at SNRs may be too efficient. 
More than 40\% of the explosion energy ($E_o$) is tranferred to 
CRs and the source CR spectrum, $N(E)\propto E^{-\alpha}$ with $\alpha \approx 1.5$, 
is too flat to be consistent with the observed CR spectrum at Earth (Ave \etal 2009).
In these models with efficient injection and acceleration,
the flow structure is significantly modified with
$\rho_2/\rho_0 \approx$ 7.2-7.5 for WA/WB models.

In the intermediate temperature ISM models ($T_0=10^5$K, $n_H=0.03{\rm cm^{-3}}$),
the flow structure is still significantly modified with
$\rho_2/\rho_0 \approx$ 5.7-6.0 
and the fraction of energy conversion, $E_{\rm CR}/E_0 \approx 0.2-0.4$ for MA/MB models.

Only in the hot ISM model ($T_0=10^6$K, $n_H=0.003{\rm cm^{-3}}$)
with inefficient injection ($\epsilon_B=0.2$ or $R_{\rm inj}>3.8$),
the shock structure is almost test-particle like with $\rho_2/\rho_0 \approx$ 4.2-4.4 
and the fraction of energy conversion, $E_{\rm CR}/E_0 \approx 0.1-0.2$ for HA/HB models.
The predicted source spectrum $G(p)$ has a slope $q=4.1-4.3$ for $10^{11}< E< 10^{15}$ eV.
Here drift of Alfv\'en waves relative to the bulk flow upstream of the subshock
plays an important role, since the modified test-particle slope, 
$q_{\rm tp}=3(u_0-v_A)/(u_0-v_A-u_2)$, can be steeper than the canonical value of $q=4$
for strong unmodified shocks. 
With magnetic fields of $B_0=30\mu$G, the Alfv\'en speed is $v_A\approx 1000 {\rm km s^{-1}}$,
and so the modified test-particle slope is $\alpha \approx 2.3$. 
This may imply that SN exploding into the hot ISM are the dominant sources
of Galactic CRs below $10^{15}$eV.
One might ask if the magnetic field amplification would take 
place in the case of such inefficient acceleration, 
since the magnetic field energy density is expected to be 
proportional to the CR pressure.  
An alternative way to enhance the downstream magnetic field
was suggested by Giacalone \& Jokipii (2007).
They showed that the density fluctuations pre-existing
upstream can warp the shock front and vortices are generated
behind the curved shock surface.  Then vortices are cascade
into turbulence which amplifies magnetic fields via turbulence
dynamo. 

Finally, in all models considered in this study, for Bohm-like diffusion with the amplified magnetic 
field in the precursor, indicated by X-ray observations of young SNRs, 
the particles could be accelerated to $E_{\rm max} \approx 10^{15.5}Z$eV.
The drift and dissipation of {\it faster} Alfv\'en waves in the precursor, on the other hand,
soften the CR spectrum and reduce the CR acceleration efficiency. 

\acknowledgments{
The author would like to thank T. W. Jones and P. Edmon for helpful comments on
the paper and Kavli Institute for Theoretical Physics (KITP) for their 
hospitality and support, where some parts of this work were carried out
during {\it Particle Acceleration in Astrophysical Plasmas 2009} program. 
This work was supported by National Research Foundation of Korea Grant 
funded by the Korean Government (2009-0075060).
}


\begin{references}

\reference{} Abdo, A. A. for the Fermi LAT collaboration, 2009
Fermi LAT Discovery of Extended Gamma-Ray Emission in the Direction of Supernova Remnant W51C,
ApJL, 706, L1

\reference{} Abdo, A. A. for the Fermi LAT collaboration, 2010
Fermi-Lat Discovery of GeV Gamma-Ray Emission from the Young Supernova Remnant Cassiopeia A,
ApJL, 710, L92 

\reference{} Aharonian, H. for the H.E.S.S. collaboration, 2004, 
High-energy particle acceleration in the shell of a supernova remnant,
Nature, 432, 75

\reference{} Aharonian, H. for the H.E.S.S. collaboration, 2009, 
Discovery of Gamma-Ray Emission From the Shell-Type Supernova Remnant RCW 86 With Hess,
ApJ, 692, 1500

\reference{} Amato, E., Blasi, P., 2006,
Non-linear particle acceleration at non-relativistic shock waves in the presence 
of self-generated turbulence,
MNRAS, 371, 1251

\reference{} Ave, M., Boyle, P. J., H\"oppner, C., Marshall, J., \& M\"uller, D., 2009,
Propagation and source energy spectral of cosmic ray nuclei at high energies,
ApJ, 697, 106

\reference{} Bamba, A., Yamazaki, R, Ueno, M. \& Koyama, K., 2003,
Small-Scale Structure of the SN 1006 Shock with Chandra Observations,
ApJ, 589, 827

\reference{} Bamba, A., Yamazaki, R, Yoshida, T., Terasawa, T., \& Koyama, K., 2006,
Small-scale structure of non-thermal X-rays in historical SNRs,
Advances in Space Research, 37, 1439 

\reference{} Bell, A.~R., 1978, 
The acceleration of cosmic rays in shock fronts. I,
MNRAS, 182, 147

\reference{} Bell, A.~R., 2004, 
Turbulent amplification of magnetic field and diffusive shock acceleration of cosmic rays,
MNRAS, 353, 550 

\reference{} Berezhko, E.~G., \& V\"olk, H.~J., 1997,
Kinetic theory of cosmic rays and gamma rays in supernova remnants. 
I. Uniform interstellar medium,
Astropart. Phys. 7, 183 

\reference{} Berezhko, E.~G., Ksenofontov, L.~T., \& V\"olk, H.~J., 2003,
Confirmation of strong magnetic field amplification and nuclear cosmic ray 
acceleration in SN 1006,
A\&Ap, 423, L11 

\reference{} Berezhko, E.~G., \& V\"olk, H.~J., 2006,
Theory of cosmic ray production in the supernova remnant RX J1713.7-3946,
A\&Ap, 451, 981 

\reference{} Berezhko, E.~G., Ksenofontov, L.~T., \& V\"olk, H. J, 2009,
Cosmic ray acceleration parameters from multi-wavelength observations. The case of SN 1006,
A\&Ap, 505, 169

\reference{} Blandford, R.~D., \& Eichler, D., 1987,
Particle acceleration at astrophysical shocks - a theory of cosmic-ray origin,
Phys. Rept., 154, 1

\reference{} Blasi, P., Gabici, S., \& Vannoni, G, 2005,
On the role of injection in kinetic approaches to non-linear particle acceleration at non-relativistic shock waves,
MNRAS, 361, 907

\reference{} Caprioli, D., Amato, E., Blasi, P., 2009,
The contribution of supernova remnants to the galactic cosmic ray spectrum,
preprint arXiv:0912.2964


\reference{} Drury, L.~O'C., Ellison, D.~E., Aharonian, F.~A. \etal, 2001, 
Test of galactic cosmic-ray source models - Working Group Report,
Space Science Reviews, 99, 329

\reference{} Giacalone, J., 2005,
The Efficient Acceleration of Thermal Protons by Perpendicular Shocks
ApJ, 628, L37

\reference{} Giacalone, J., Jokipii, J. R., 2007,
Magnetic Field Amplification by Shocks in Turbulent Fluids,
ApJ, 663, L41

\reference{} Helder, E.~A. et al., 2009,
Measuring the Cosmic-Ray Acceleration Efficiency of a Supernova Remnant,
Science, 325, 719

\reference{} Jones, T.~W., 1993, 
Alfv\'en wave transport effects in the time evolution of parallel cosmic-ray-modified shocks
ApJ, 413, 619

\reference{} Kang, H., 2006,
Cosmic ray acceleration at blast waves from Type Ia Supernovae,
Journal of Korean Astronomical Society, 39, 95 (Paper I)

\reference{} Kang, H., \& Jones, T.~W., 1995,
Diffusive Shock Acceleration Simulations: Comparison with Particle Methods and Bow Shock Measurements,
ApJ, 447, 944

\reference{} Kang, H., Jones, T.~W., LeVeque, R.~J., \& Shyue, K.~M., 2001,
Time Evolution of Cosmic-Ray Modified Plane Shocks, ApJ, 550, 737

\reference{} Kang, H., Jones, T.~W., \& Gieseler, U.~D.~J., 2002, 
Numerical Studies of Cosmic-Ray Injection and Acceleration, 
ApJ, 579, 337

\reference{} Kang, H., \& Jones, T.~W., 2006,
Numerical studies of diffusive shock acceleration at spherical shocks,
Astropart. Phys. 25, 246

\reference{} Kang, H., \& Jones, T.~W., 2007,
Self-similar evolution of cosmic-ray-modified quasi-parallel plane shocks,
Astropart. Phys. 28, 232

\reference{} Kang, H., Ryu, D., \& Jones, T.~W., 2009,
Self-Similar Evolution of Cosmic-Ray Modified Shocks: The Cosmic-Ray Spectrum,
ApJ, 695, 1273

\reference{} Koyama, K., Petre, R., Gotthelf, E.~V. \etal, 1995,
Evidence for Shock Acceleration of High-Energy Electrons in the Supernova Remnant SN:1006, 
Nature, 378, 255

\reference{} Lagage, P.~O., \& Cesarsky, C.~J., 1983,
The maximum energy of cosmic rays accelerated by supernova shocks,
A\&Ap, 118, 223

\reference{} Lucek, S.~G., \& Bell, A.~R., 2000,
Non-linear amplification of a magnetic field driven by cosmic ray streaming, MNRAS, 314, 65

\reference{} Malkov, M.~A., \& Drury, L.~O'C., 2001,
Nonlinear theory of diffusive acceleration of particles by shock waves,
Rep. Progr. Phys. 64, 429

\reference{} Malkov, M.~A., \& V\"olk, H.~J., 1998,
Diffusive ion acceleration at shocks: the problem of injection,
Adv. Space Res. 21, 551

\reference{} Morlino, G., Amato, E., \& Blasi, P., 2009,
Gamma-ray emission from SNR RX J1713.7-3946 and the origin of galactic cosmic rays,
MNRAS, 392, 240

\reference{} Parizot, E., Marcowith, A., Ballet, J., \& Gallant, Y. A., 2006,
Observational constraints on energetic particle diffusion in young supernovae remnants: 
amplified magnetic field and maximum energy
A\&Ap, 453, 387

\reference{} Reynolds, S.~P., 2008,
Supernova Remnants at High Energy,
Annu. Rev. of Astro. Astrophys., 46, 89


\reference{} Skilling, J., 1975,
Cosmic ray streaming. I - Effect of Alfv\'en waves on particles,
MNRAS, 172, 557

\reference{} Spitzer, L.~J., 1978, 
{\em Physical Processes in the Interstellar Medium} 
(John Wiley and Sons, New York).

\reference{} V\"olk, H.~J., Berezhko, E.~G., \& Ksenofontov, L.~T., 2003,
Variation of cosmic ray injection across supernova shocks,
A\&Ap, 409, 563

\reference{} V\"olk, H.~J., Berezhko, E.~G., \& Ksenofontov, L.~T., 2005,
Magnetic field amplification in Tycho and other shell-type supernova remnants
A\&Ap, 433, 229 

\reference{} Vladimirov, A.~E., Bykov, A.~M., Ellison, D.~C., 2008,
Turbulence Dissipation and Particle Injection in Nonlinear Diffusive 
Shock Acceleration with Magnetic Field Amplification,
ApJ, 688, 1084

\reference{} Zirakashvili, V. N., \& Ptuskin, V. S., 2008,
The influence of the Alfv\'enic drift on the shape of cosmic ray spectra in SNRs,
{\it Proceedings of the 4th International Meeting on High Energy Gamma-Ray Astronomy},
AIP Conference Proceedings, 1085, 336
\end{references}
\end{document}